\documentclass[10pt,twocolumn,twoside,submit]{JCNtran}

\usepackage[dvips]{epsfig}
\usepackage[latin1]{inputenc}
\usepackage[T1]{fontenc}

\usepackage{subfigure}

\usepackage{graphicx}
\usepackage{epstopdf}
\DeclareGraphicsExtensions{.eps}

\begin{document}
\bibliographystyle{jcn}

\title{A Proxy Acknowledgement Mechanism for \\TCP Variants in Mobile Ad Hoc Networks}
\author{May Zin Oo, Mazliza Othman and Timothy O'Farrell 
\thanks{Manuscript received January 23, 2014; approved for publication by Grace Kim, October 1, 2015.}
\thanks{May Zin Oo is with the Department of Information Technology Engineering, Mandalay Technological University, Myanmar, email: mayzinoo.mz@gmail.com. } \thanks{Prof. Mazliza Othman is with the Faculty of Computer Science and IT, University of Malaya, Malaysia, email: mazliza@um.edu.my. } \thanks{Prof. Timothy O'Farrell, Chair of Wireless Communications is with department of Electronic \& Electrical Engineering, University of Sheffield, UK, email:t.ofarrell@sheffield.ac.uk. } }
\maketitle

\begin{abstract}
A sequence number checking technique is proposed to improve the performance of TCP connections in mobile ad hoc networks. While a TCP connection is initialized, a routing protocol takes the responsibility for checking the hop count between a source and destination pair. If the hop count is greater than a predefined value, the routing protocol decides to use a proxy node. The responsibility of a proxy node is to check the correctness of data packets and inform the missing packets by sending an acknowledgement from a proxy node to the source node. By doing so, the source node is able to retransmit any missing packet in advance without waiting until an end-to-end acknowledgement is received from the destination. Simulation results show that the proposed mechanism is able to increase throughput up to 55\% in static network and decrease routing overhead up to 95\% in mobile network.
\end{abstract}

\begin{keywords}
	MANET, TCP variants, proxy-assisted routing, proxy acknowledgement.
\end{keywords}

\section{\uppercase{Introduction}}
\label{sec:introd}
\PARstart{T}{ransmission} control protocol (TCP) ~[1] is an independent protocol that is not related to the underlying network technology.TCP's reliability depends on the retransmission of lost packets. Packet losses occur frequently as the amount of traffic to access Internet applications increases. As there are no resource reservation and admission control to monitor the imposed network load, the total number of packets is more than they should be, leading to congestion. The TCP congestion control mechanism controls the sending rate by keeping track of the congestion window (CWND). When a new TCP connection is established, the CWND is set to one maximum segment size (MMS) ~[2]. When an acknowledgement packet is received, the CWND is increased exponentially. To limit the CWND size, there is a slow-start threshold that is set to 65Kbytes ~[3]. After the CWND size has reached the slow start threshold, it is increased linearly. This region is called congestion avoidance. In the congestion avoidance region, the initial window is increased linearly. Whenever the timeout occurs, the slow start procedure is initialized by setting the CWND to one, whereas the CWND is halved (i.e., CWND = CWND/2) when three duplicate ACKs are received. Initiating the slow start procedure whenever a timeout event occurs reduces throughput.

TCP controls network congestion and bridges a file transfer application. TCP limits the transmission rate depending on the congestion window and provides reliability using the packet retransmission technique. It sends ack packets for each received TCP segment. Sending ack packets for all received TCP segments reduce the performance throughput, especially in longer path length. Therefore, ~[4] not only mentions that the path length is an important factor to consider when choosing appropriate delay window sizes, but also proposes a delay acknowledgement scheme that delays and balances ack flow and burst loss. In addition, a number of TCP variants have been introduced for the traditional TCP (TCP-Tahoe) with layer approach (i.e., enhance within the single layer) and cross layer approach (i.e., use the aid of other layers). The following TCP variants are layer approach enhancements.

\textit{TCP Reno:} Instead of starting transmission from a slow start after a relatively long idle period, ~[5] introduced TCP-Reno that adds fast retransmit and fast recovery algorithms. With fast retransmit, Reno attempts to retransmit packets before a timeout, but the sender initiates the slow-start procedure as if a timeout causes the retransmission. With fast recovery, Reno uses additive increase/multiplicative decrease at all time, and only initiates the slow start when either a connection is established or a timeout occurs. In other words, Reno with fast recovery omits the slow start if no timeout occurs.

\textit{TCP-NewReno:} NewReno ~[6] is an improvement of Reno, and it is advanced with fast transmit, where three duplicate acknowledgments signal a retransmission without a timeout with fast recovery. The fast recovery means that once a certain ack threshold is received, the window size is decreased by half rather than starting over with a slow start. New Reno increases the adoption of the TCP selective acknowledgement (SACK) modification ~[7]. It involves two kinds of acks: partial ack and full ack. The partial ack acknowledges some segments at the fast recovery stage while the full ack acknowledges all outstanding data. Upon receiving the full ack, the sender sets the congestion window to slow start threshold and terminates the fast recovery. Then the congestion avoidance mechanism is resumed. In this way, the New Reno maintains a high throughput.

\textit{TCP Vegas:} Reno and NewReno variants are window-based transport protocols that adjust the congestion window upon packet losses. On the other hand, ~[8] introduces a delay-based TCP, called TCP-Vegas, which does not violate the congestion avoidance paradigm. Vegas prevents packet losses by reducing the sending rate once it senses initial congestion. Vegas uses packet delay as an indication of congestion. In a situation where a duplicate ack is received, the timestamp for the ack is compared to a timeout value. If the timestamp is greater than the timeout value, then Vegas will retransmit rather than waiting for three duplicate acks. Vegas estimates the available bandwidth using the difference between expected and actual flow rates. When the network is congested, the actual flow rate will be smaller than the expected flow rate. Otherwise, the actual flow rate and the expected flow rate are close to each other. TCP-Vegas estimates the congestion level and updates the window size accordingly.

\textit{TCP Westwood:} TCP-Westwood ~[9] modifies the congestion window algorithm of TCP at a sender-site. The idea behind is to estimate the available bandwidth to control the congestion window size by monitoring the ack packets. A sender measures the rate of acks that it receives and estimates the current bandwidth according to that connection. Once packet losses occur (i.e. timeout or duplicate acks), the sender sets appropriate congestion window according to the estimated bandwidth. Instead of halving the congestion window like New Reno, Westwood backs off some value of cwnd and threshold based on the estimated value to ensure faster recovery. The improvement of Westwood is more significant in lossy wireless link due to its efficient bandwidth estimation.

After describing the variants of TCP, the rest of this paper is organized as follows. Section II describes the problem description of why the end-to-end checking of TCP is insufficient in the mobile networks. Section III presents the proposed solution that can transmit data efficiently. Section IV presents the simulation results depicting performance analysis by exploiting Grid and Random topologies. Finally, section V concludes the paper.

\section{\uppercase{Problem Description}}
Recently, there is an approach ~[10] that considers TCP mobility solution to enable seamless session end-point migration, and builds on a lightweight MPTCP proxy function. For the reliability of TCP data packets at the transport layer, the sender and receiver checks whether the packets are received correctly by using acknowledgment packets. However, not only are the routes very unstable, but also the end-to-end data reliability checking of TCP is insufficient in the mobile ad hoc networks. The hop-by-hop transport protocols proposed by ~[11, 12] checks the reliability of data packet at every node to improve the correctness of data and throughput. However, this approach is very expensive on scarce resources, such as bandwidth, power and memory usages. Split TCP approach ~[13] also splits a TCP connection into a set of shorter TCP connections, and proxy nodes are used as end points of these short connections. The number of proxies depends on the path length. This approach works best with fewer TCP connections. However, as the number of TCP connections increases, not only the number of proxies, but also the overhead for these proxies increases. A proxy acknowledgement (PACK) mechanism is thus proposed for TCP senders, in which only a proxy node is used for each TCP connection. Instead of checking TCP packets either at every hop or at many proxies like split TCP, the proposed mechanism monitors packets at one proxy node between a source and destination pair while maintaining the end-to-end reliability of TCP. The objective of this research is to address end-to-end and hop-by-hop inefficiency at the transport layer by using a proxy-assisted approach.

\section{\uppercase{Problem Solution}}
The proposed mechanism must satisfy the following requirements to transmit data efficiently.\\ 
\textit{Reliable delivery:} This mechanism ensures the delivery of all application-layer data packets to the destination node with the aid of a proxy node. Hence, the protocol guarantees the reliability of data packets and throughput improvement by detecting the missing sequence numbers and acknowledging them to the source node in advance.\\
\textit{Interoperability with the application layer:} The mechanism does not affect the end-to-end interface for reliable transmission of TCP packets.\\
\textit{Cross-layer information:} The protocol requires more than a minimum of cross-layer information. In particular, it needs information, such as the MAC address of upstream nodes for the unicast transmission; proxy selection from the routing layer to detect the missing TCP sequence numbers and number of missing sequence numbers; and monitoring ACK packets from the TCP destination at the routing layer to inform the missing sequence number to the source node.\\
\textit{Acknowledgement Scheme of a Proxy Node:} In the proxy node acknowledgement scheme (PACK), the proxy node always monitors the TCP sequence number, records missing sequence numbers, and informs these to the source node. In this way, the source node can retransmit the missing sequence numbers in advance of the end-to-end acknowledgements. The following are the research questions that must be addressed to do so.

\begin{enumerate}
	\item How can a proxy node know which TCP sequence numbers are missing?
	\item How does a proxy node inform the missing sequence numbers to the source? 
	\item What is the function of the intermediate nodes upon detecting of the missing numbers? 
	\item How does the source node retransmit the missing sequence number before the end-to-end ACK is received?  
\end{enumerate}
\subsection{Proxy Selection}
 The proxy selection process is the function of routing layer protocol. As mentioned in our paper ~[14], a middle node is selected as a proxy node according to the path length or hop count (HC). When a destination node receives a route request packet (RREQ), it decides whether to use the assistance of a proxy node by checking the HC of RREQ. In this case, the minimum allowable hop count value, which is an adjustable value, is defined as three. When the distance between the source and destination node is less than the defined value, a proxy node is not used. If the hop count of RREQ is greater than the predefined value, the proxy calculation function is invoked. After calculating a Proxy HC (PHC) value, the destination node attaches this value in the RREP packet to assign a proxy node. At this point, the destination node only knows how many hops away the proxy node is, but does not know the proxy address yet. The intermediate nodes that receive the RREP packet compare the calculated PHC value and hop count value of RREP packet, which increases monotonically. If these values are equal, this node is assigned to perform as a proxy node. Otherwise, RREP packet is forwarded to the next hop. Proxy-assistance routing (PART) and its detail routing mechanism are referred to our paper ~[14].
\subsection{Proxy Failure Condition}
The proxy node has to be intelligent to detect whether it should give up the proxy duty. Whenever an ERROR message is received, the proxy node tries to repair the route first. However, when the proxy encounters more than three ERRORs continuously within a very short time, it assumes that it is out of the transmission range or has moved far away from the source and destination, and sends ERROR with a RESET flag to the source node. On the other hand, there could be a situation that the proxy node moves out of the transmission range and is unavailable to inform the source node. In this case, the source node uses the timer and retry techniques to discover a new route if it does not receive the ERROR signal from the proxy node. The best thing the proxy node can do is to repair a route with less overhead and supports data transmission with less delay.

The responsibility of a proxy node for a TCP connection is to detect the missing TCP sequence number (\textit{miss\_seqno}) and the number of missing sequence numbers (\textit{num\_miss\_seqno}) by monitoring the data packets going through the routing layer of a proxy node and send a PACK to the source node to inform it of \textit{miss\_seqno}. To detect the \textit{miss\_seqno}, a proxy node uses a simple algorithm to check the current sequence number (\textit{cur\_seqno}), which is in the TCP packet header, and the expected sequence number (\textit{exp\_seqno}), which is increased linearly. Fig. 1 and Fig. 2 show the sequence number checking algorithm. While a source node is sending TCP data packets to the destination, a proxy node monitors \textit{seqno} and checks for \textit{miss\_seqno}. Then, the proxy node puts the sequence number of the TCP header in the \textit{cur\_seqno}.

\subsection{Sequence Number Checking at a Proxy Node}
The \textit{seqno} always starts at one. In order to increase \textit{exp\_seqno} linearly, the \textit{cur\_seqno} and \textit{exp\_seqno} must initially be the same. However, if the proxy node change occurs, these \textit{seqno} may not start from one. Therefore, the value of \textit{exp\_seqno} must be updated according to the \textit{cur\_seqno}. To detect the proxy change, the current proxy node ID is recorded in the routing table of each node. In other words, when the proxy node detects that \textit{cur\_seqno} is equal to one (i.e., the foremost seqno), this proxy node's id is added in the \textit{now\_proxy field} in the routing table (line 6 in Fig. 2). Later, if the current proxy is not equal to the \textit{now\_proxy}, a proxy change event has occurred. The \textit{now\_proxy} field of the routing table is then updated with a new proxy (line 10 in Fig. 2), and \textit{exp\_seqno} is updated with \textit{cur\_seqno}. While the TCP seqno is monitored, \textit{exp\_seqno} is increased by one for every received seqno, and then the \textit{cur\_seqno} and \textit{exp\_seqno} are compared. In Fig. 1, as long as the \textit{cur\_seqno} and \textit{exp\_seqno} are the same, the proxy node assumes that there is no missing sequence number. After receiving seqno 4, seqno 5 is supposed to receive in \textit{cur\_seqno} according to the exp\_seqno. However, seqno 7 is received in the cur\_seqno. In other words, \textit{cur\_seqno} is greater than \textit{exp\_seqno}, which indicates a missing sequence number. Therefore, the proxy node adds \textit{seqno 5} in the missing sequence number and calculates the number of missing sequence numbers by subtracting \textit{cur\_seqno} and \textit{exp\_seqno}. Then, the \textit{exp\_seqno} is updated by adding \textit{cur\_seqno} value, for example, \textit{exp\_seqno 7} in Fig. 1.

\begin{figure}[h!]
	\centering
\includegraphics[scale=0.46]{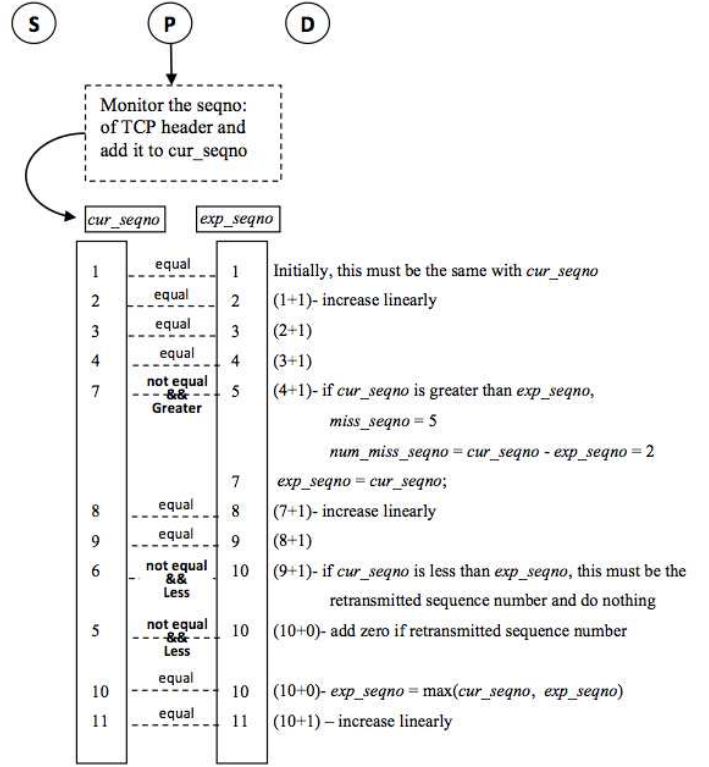}
\caption{Sequence number checking algorithm}
\label{fig:Fig1}
\end{figure}

\begin{figure}[h!]
	\centering
\includegraphics[scale=0.41]{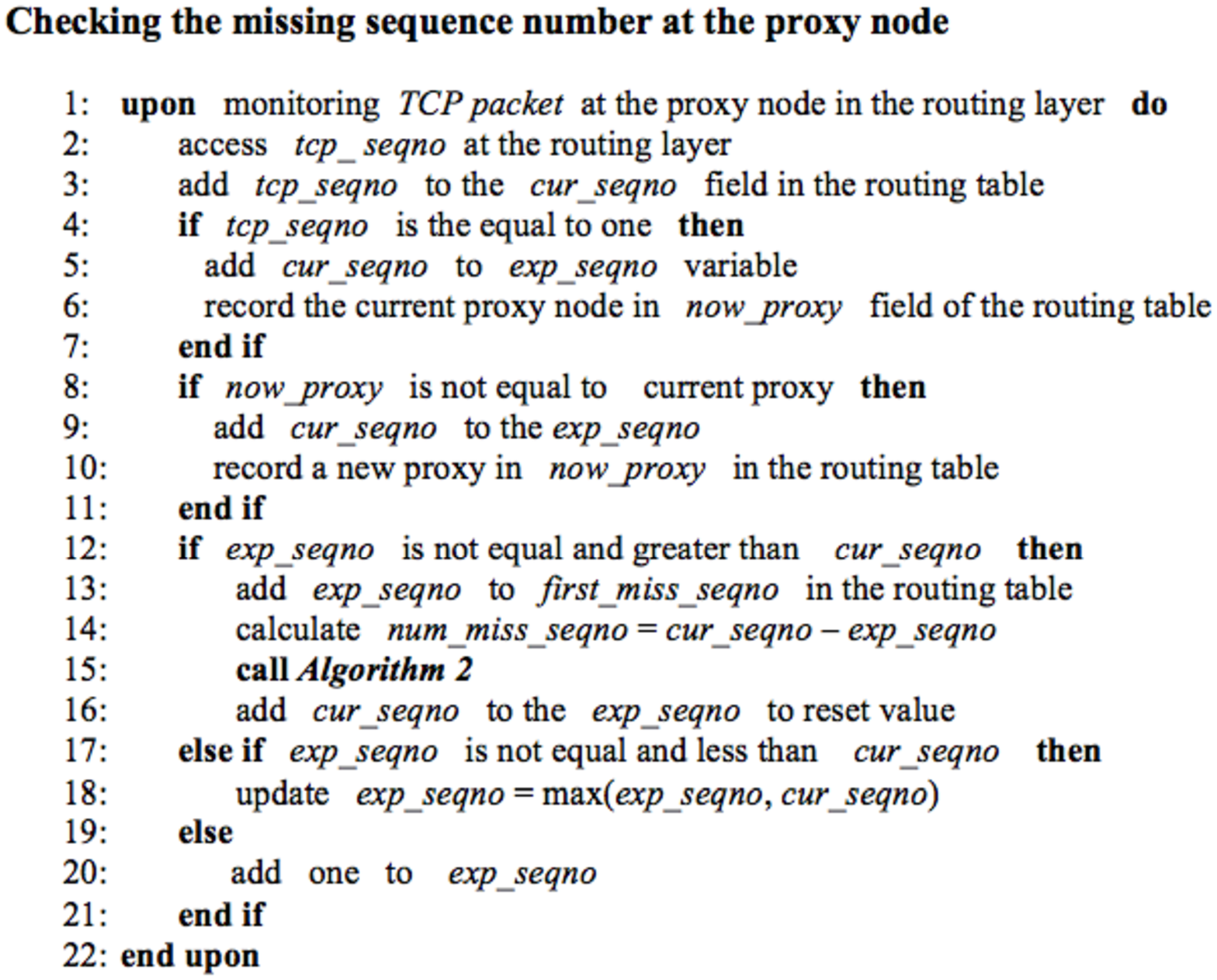}
\caption{Sequence number checking at a proxy}
\label{fig:algo1}
\end{figure}

After receiving \textit{seqno 9}, the proxy node detects that the \textit{cur\_seqno} and \textit{exp\_seqno} are not equal, and the \textit{cur\_seqno} is less than \textit{exp\_seqno}, meaning that the retransmitted \textit{miss\_seqno} is received. Instead of increasing \textit{exp\_seqno} linearly, it is added to zero as shown in Fig. 1. To update \textit{exp\_seqno} for checking the rest of the seqno, the proxy node takes the maximum between \textit{cur\_seqno} and \textit{exp\_seqno} as shown in line 18 of Fig. 2. As soon as the proxy node detects that the sequence numbers are missing, it sends a PACK packet by adding \textit{miss\_seqno} and \textit{num\_miss\_seqno} to inform the source node. The proposed algorithm informs to the source node in advance before an end-to-end acknowledgement. 

\subsection{Cross layer information for unicast PACK}
The PACK packet is sent by unicast towards the source node by using the MAC layer information. When a PACK is received, the intermediate nodes are responsible for checking whether it is the source. If it is, it updates its routing table with the \textit{miss\_seqno}. Otherwise, it updates the routing table and forwards the packet to the next hop towards the source node. Fig. 3 shows the procedure for sending PACK. Before a proxy node sends PACK, it looks at the MAC layer address of the upstream nodes, adds it in the \textit{next\_hop} field of the routing table, and sends PACK with the \textit{miss-seqno} information to the source.

\begin{figure}[h!]
\centering
\includegraphics[scale=0.37]{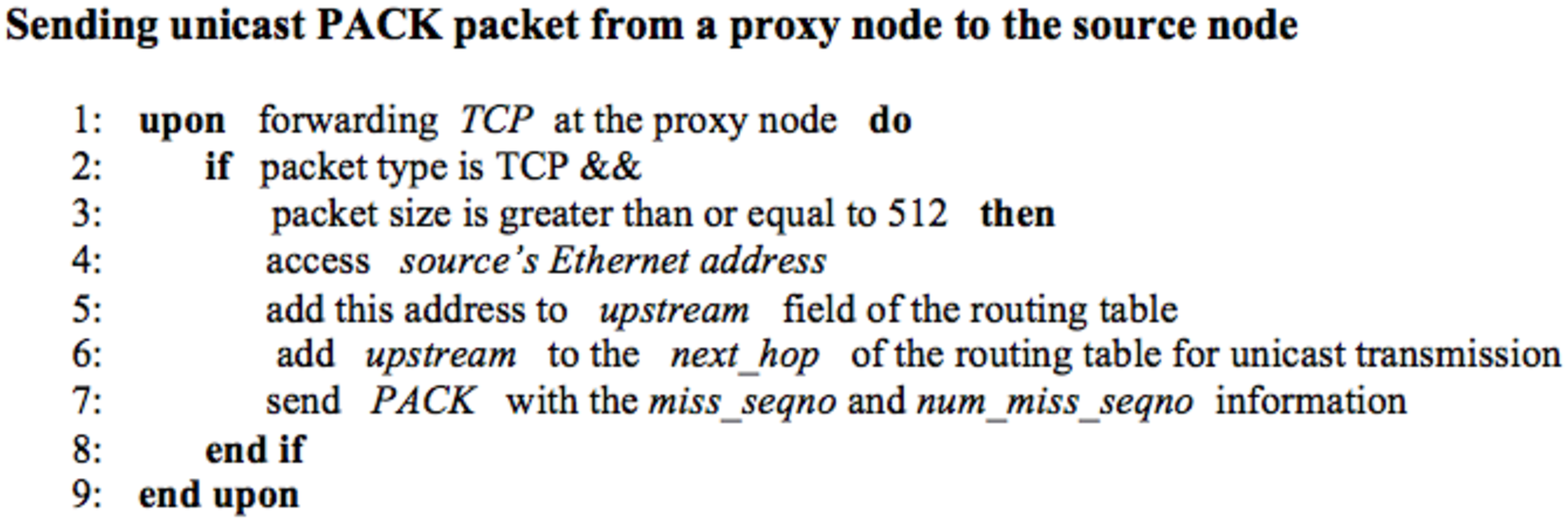}
\caption{Sequence number checking at a proxy}
\label{fig:algo2}
\end{figure}

\subsection{One-hop broadcast}
Proxy node broadcasts one hop broadcast (OHPACK) packet so that the neighboring nodes of a proxy can update the \textit{miss\_seqno} correctly. When the neighbors receive the OHPACK, they update their routing tables by adding \textit{miss\_seqno} and \textit{num\_miss\_seqno}. As there are many possibilities of ack paths, PACK and OHPACK packets are used to attach \textit{miss\_seqno} and \textit{num\_miss\_seqno} values in the ack packets to inform the source node.

\subsection{Monitoring ack packet from the destination}
An important problem to address is how to inform the \textit{miss\_seqno} to the transport layer of source node. Because PACK and OHPACK are lower routing layer packets, it is impossible to send \textit{miss\_seqno} to the upper transport layer. Moreover, mentioned packet transmission functions work locally at the routing layer. Therefore, there is only one way to modify the end-to-end ack packet, which is by adding new fields in the TCP packet header. When the end-to-end ack packets are received, the responsibilities of the intermediate nodes are to check whether it is the next hop towards the source node. If it is, the node adds the \textit{miss\_seqno} and \textit{num\_miss\_seqno} values to the end-to-end ack packets to inform the source which TCP sequence numbers are missing. Otherwise, the node simply forwards the ack packets to its neighboring nodes in the network.

\subsection{Functions of TCP data source}
The TCP data source has to check every received ack packet. When the \textit{miss\_seqno} value is not equal to zero, the source retransmits the \textit{miss\_seqno}, and decides the number of missing sequences according to the \textit{num\_miss\_seqno}. If the \textit{num\_miss\_seqno} is equal to one, TCP source adds the \textit{miss\_seqno} in the TCP's header and sets the \textit{retransmit\_timer}. Then, the TCP packets are sent in advance before receiving the end-to-end acks for those packets. Otherwise, the TCP source makes sure to send all missing sequences by counting the \textit{num\_miss\_seqno}. We apply the PACK mechanism to the TCP's variants and implement them in network simulator. Then, we analyze the performance of each transport protocol with PART and AODV in static and mobile environments. Our previously proposed PART routing protocol is used because PACK mechanism is built on it.

\section{\uppercase{Experimental Analysis}}
\textit{Objective of experiment:} We analyze the performance of the PACK scheme $7×7$ grid in static network and random node movements in mobile network.\\ \textit{Expected outcome:} The sequence number checking of the PACK mechanism positively influences the best performance for TCP variants in terms of the throughput, packet loss rate and delay. 

\subsection{Grid topology in static network}
 We use Network simulator (NS2) ~[15] to carry out the simulations. Firstly, we measure the performance across the Reno, New Reno, Vegas and Westwood by using AODV and PART. To ascertain the efficiency of the PACK mechanism, we analyze its performance in a $7×7$ grid. We set the simulation area to $6000m × 2000m$. The simulations are run for 360 seconds with 5 FTP connection-flows between the sources and destinations.

\subsubsection{Throughput measurements for TCPs}
We investigate the throughput of TCP variants with PACK mechanism over PART and AODV routing protocols. Fig. 4 shows the throughput comparisons of traditional TCP (Tahoe). Tahoe-PACK over PART outperforms almost 47\% better if compared to Reno over AODV. Also in Fig. 5, New Reno-PACK offers a higher throughput almost 55\% if compared to over AODV. The PACK mechanism has the ability to retransmit the missing sequence number in advance after checking them at a proxy node.

\begin{figure}[h!]
\centering
\includegraphics[scale=0.9]{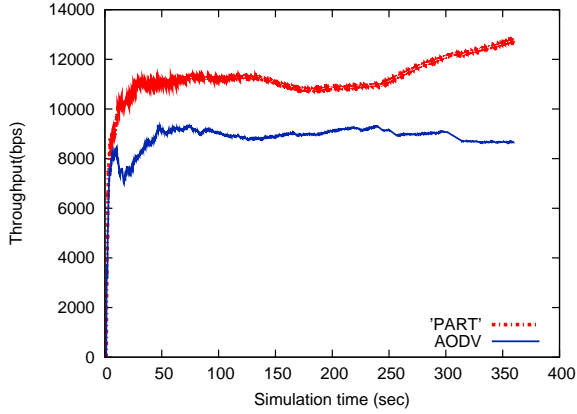}
\caption{Throughput analysis across Tahoe}
\label{fig:Tahoe_grid}
\end{figure}

\begin{figure}[H]
\centering
\includegraphics[scale=0.9]{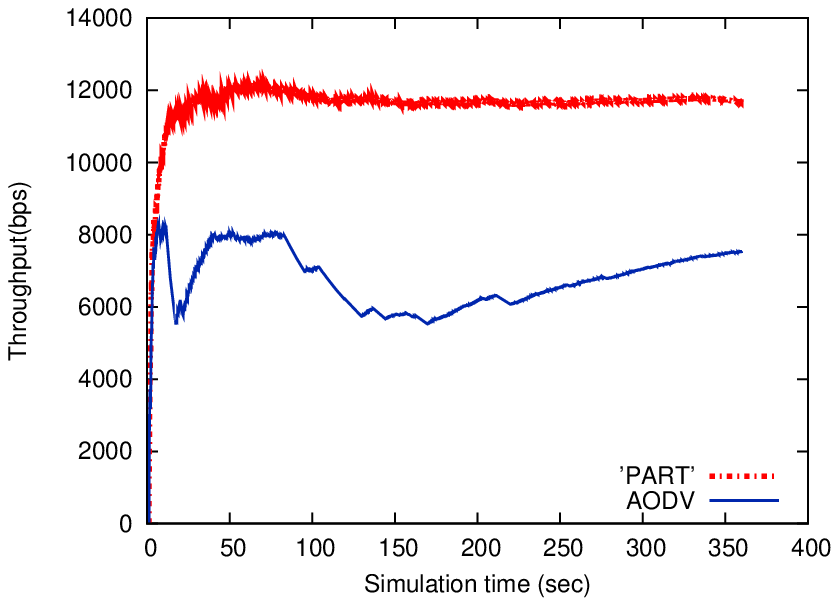}
\caption{Throughput analysis across NewReno}
\label{fig:NewReno_grid}
\end{figure}
 
\begin{figure}[h!]
\centering
\includegraphics[scale=0.9]{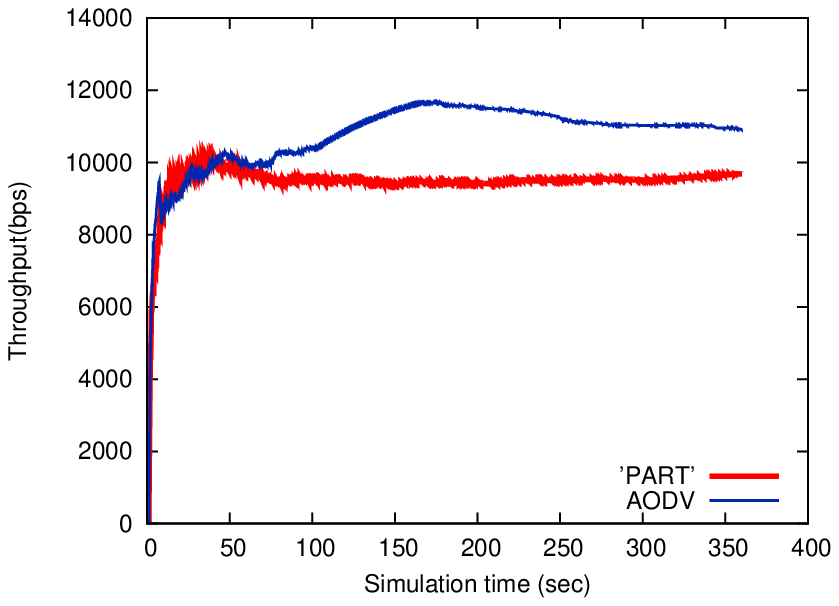}
\caption{Throughput analysis across Vegas}
\label{fig:Vegas_grid}
\end{figure}

\begin{figure}[h!]
\centering
\includegraphics[scale=0.9]{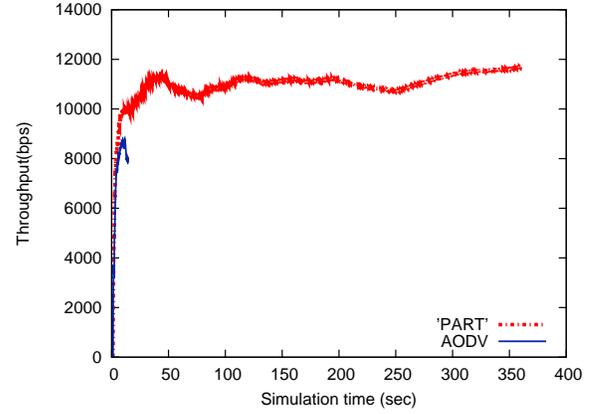}
\caption{Throughput analysis across Westwood}
\label{fig:Westwood_AODV}
\end{figure}

The performance of Vegas-PACK with PART protocol is not very good if compared to AODV as shown in Fig. 6. As it suffers throughput degradation about 12\% lower in $7×7$ grid. Although this mechanism works fine with window-based TCP protocols, its performance does not impact on the delay-based protocol like Vegas. Fig. 7 proves a better performance over Westwood. Although we analyze the performance in a static network, the route errors happen more often due to the collisions at MAC layer. As we use the fast retransmission of PACK mechanism with the PART protocol, the significant performance improvement can be seen in analysis results.

\subsubsection{Packet loss rate measurements for TCPs}
Although PACK does not significantly reduce average delay, it reduces the percentage of packet loss rates as shown in Fig. 8. Tahoe-PACK over PART reduces packet losses by almost 13\%. Also a significant achievement of New Reno-PACK is observed where it reduces packet losses by almost 46\%. The percentage of packet loss rate for Westwood-PACK over PART is 54\% lower if compared to Westwood over AODV. The PACK mechanism still affects on the performance of Vegas, in which the packet loss rates increase when compared to Vegas over AODV.

\begin{figure}[h!]
\centering
\includegraphics[scale=0.7]{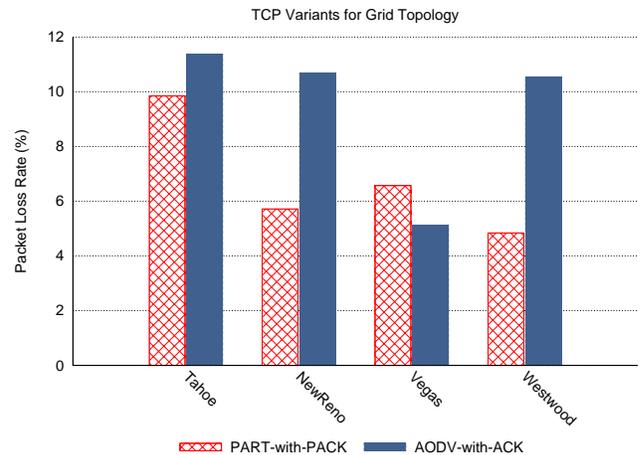}
\caption{Packet loss rate measurement across TCPs}
\label{fig:PktlossForGrid}
\end{figure}

\subsubsection{Average delay measurements for TCPs}
When we measure the average delay through $7×7$ grid, all TCP variants with PACK perform worse than AODV as shown in Fig. 9. Although PACK mechanism is able to increase throughput, its delay is a bit higher due to the extra acks from a proxy node and OHPACK packets to inform the missing sequence numbers to the intermediate nodes that are on the track of return ack to the source node.

\begin{figure}[h!]
\centering
\includegraphics[scale=0.7]{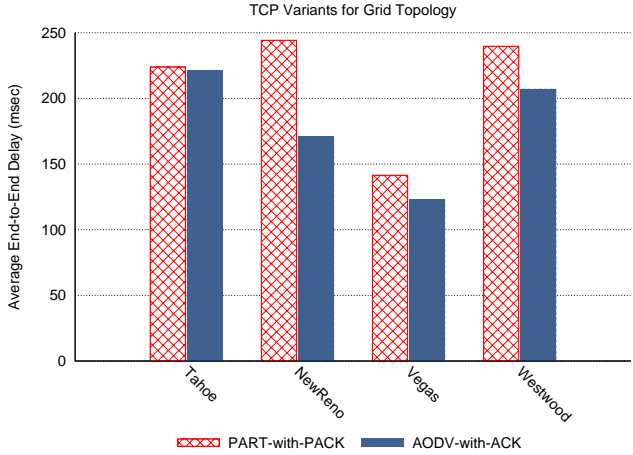}
\caption{Average delay measurement across TCPs}
\label{fig:DelayForGrid}
\end{figure}

\subsection{Random topology in mobile network}
To assure the effectiveness of the PACK mechanism, the random topologies are simulated across the mobile environment. For this purpose, we run 30 nodes in 1500m $ × $ 300m area for 360 seconds. The 10 FTP applications are exchanged between sources and destinations. Node speeds are varied with pause time 100 seconds.

\subsubsection{Throughput measurements for TCPs}
When we measure the throughput by varying node speeds during the simulation. Tahoe-PACK over PART outperforms by 3\% higher at 1m/s and 11 \% higher at 20m/s as shown in Fig. 10. As the node speed increases, the possibility of route errors increases due to the collisions and route breaks, which can be seen in detailed output trace file. However, one of the advantages of PART is that it is able to reduce the possibility of route breaks using a proxy node between every source and destination pair. Simulation results show that New Reno-PACK over PART outperforms AODV by almost 4\% higher at 20m/s node speed as shown in Fig. 11. Also in Fig. 12, Vegas-PACK over PART achieves a 11 \% higher throughput at 20m/s node speed.

\begin{figure}[!h]
	\centering
	\subfigure[Node Speed 1m/s]{\includegraphics*[scale=0.9]{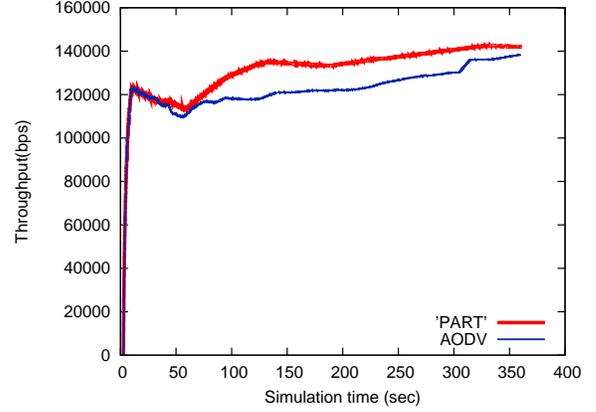}}
	\label{fig:Tahoe_1ms_random}\\
	\subfigure[Node Speed 20m/s]{\includegraphics*[scale=0.9]{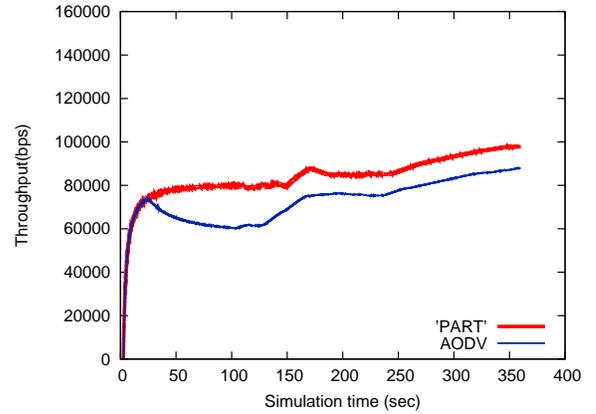}}
	\label{fig:Tahoe_20ms_random}
	\caption{Throughput analysis across Tahoe}
\end{figure}

Although we have encountered the weakness of PACK over TCP-Vegas in the static network, the significant throughput improvements are observed as the node movements and TCP traffic are randomized during the simulation in mobile ad hoc network. The authors in [9] also discovered that TCP-Westwood performs well over lossy links in wireless network. Our simulation results also show the similar observation that Westwood's performance is more significant when the node speed increases. In Fig. 13, Westwood-PACK performs 10\% higher at 10m/s and 11\% higher at 20m/s node speed if compared to end-to-end ACK over AODV.

\begin{figure}[!h]
	\centering
	\subfigure[Node Speed 1m/s]{\includegraphics*[scale=0.9]{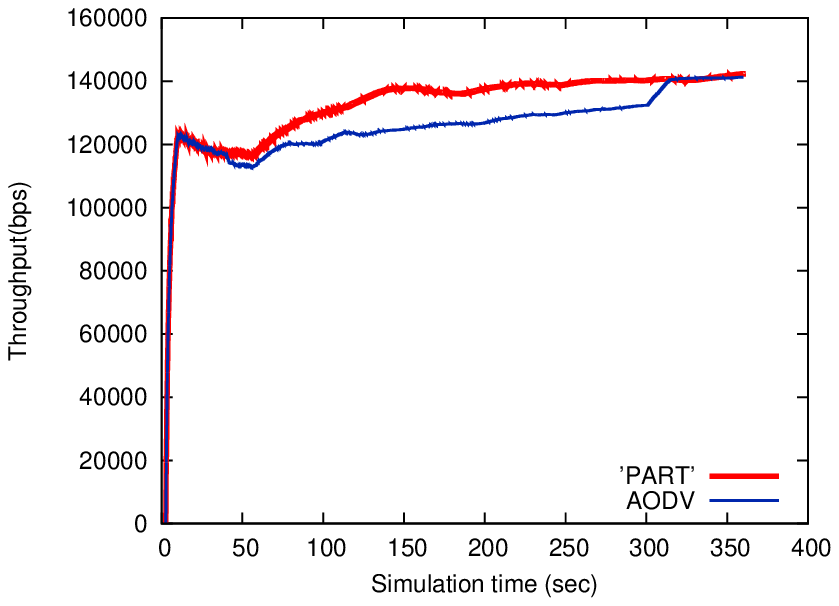}}
	\label{fig:NewReno_1ms_random}\\
	\subfigure[Node Speed 20m/s]{\includegraphics*[scale=0.9]{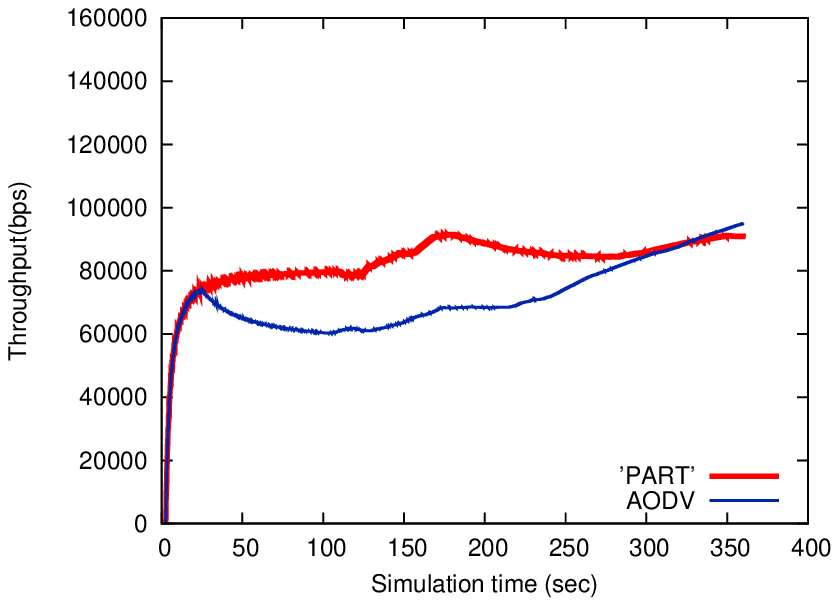}}
	\label{fig:NewReno_20ms_random}
	\caption{Throughput analysis across NewReno}
\end{figure}

\begin{figure}[!h]
	\centering
	\subfigure[Node Speed 1m/s]{\includegraphics*[scale=0.9]{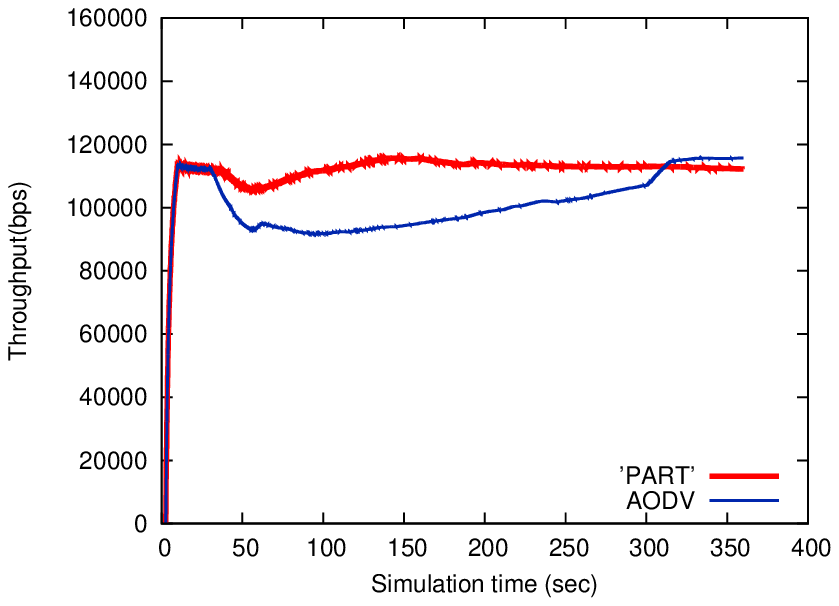}}
	\label{fig:Vegas_1ms_random}\\
	\end{figure}
	
\begin{figure}[H]
	\centering
	\subfigure[Node Speed 20m/s]{\includegraphics*[scale=0.9]{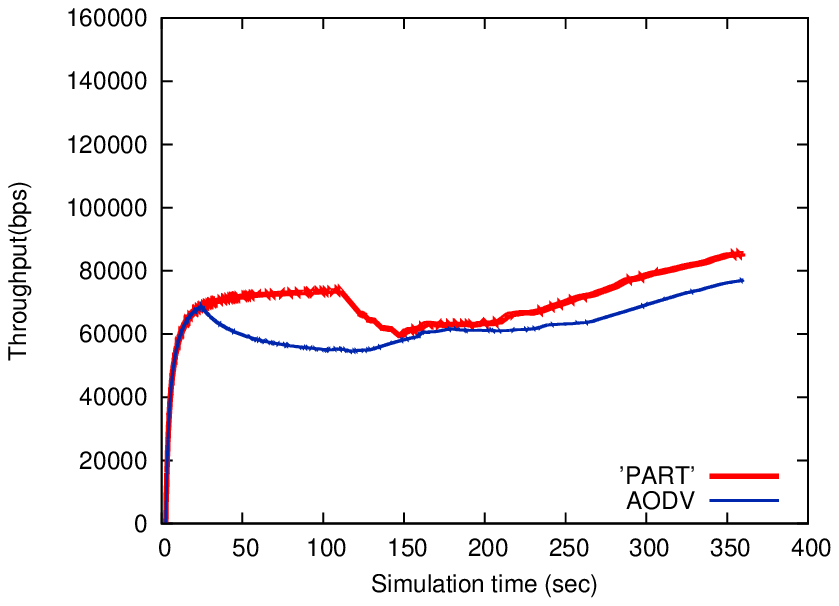}}
	\label{fig:Vegas_20ms_random}
	\caption{Throughput analysis across Vegas}
	\end{figure}

\begin{figure}[!h]
	\centering
	\subfigure[Node Speed 1m/s]{\includegraphics*[scale=0.9]{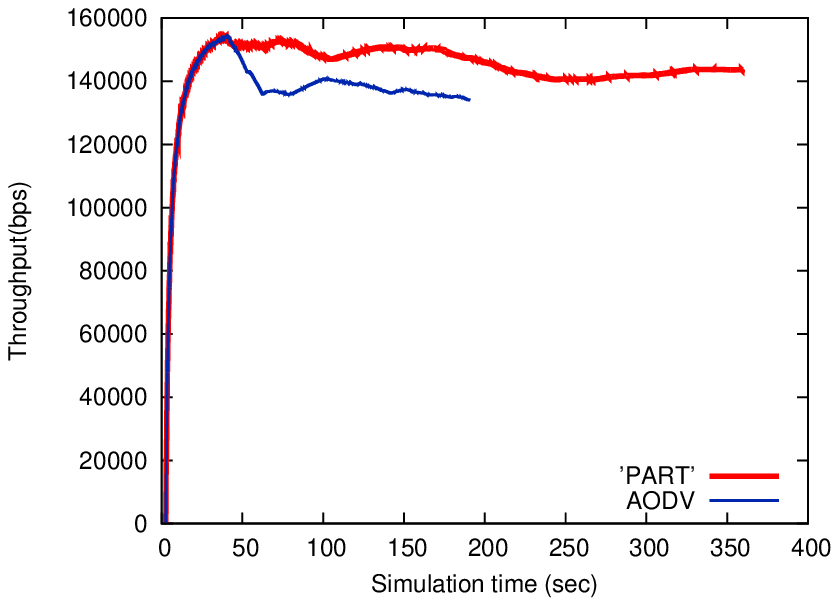}}
	\label{fig:WW_1ms_random}\\
	\subfigure[Node Speed 20m/s]{\includegraphics*[scale=0.9]{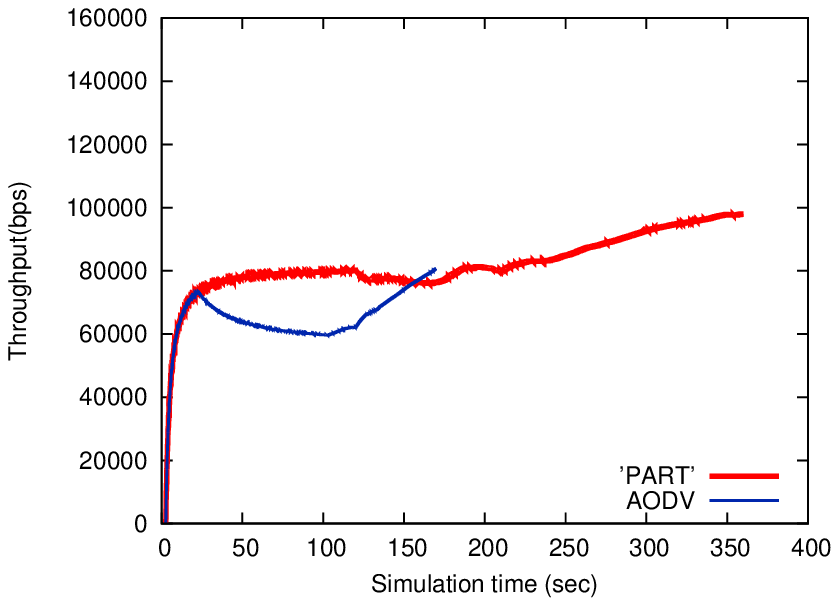}}
	\label{fig:WW_20ms_random}
	\caption{Throughput analysis across Westwood}
\end{figure}

\subsubsection{Routing overhead measurement for TCPs}
Finally, we investigate the effects of node speed on the PACK mechanism with TCP variants over PART and AODV routing protocols. As shown in Fig. 14, TCP variants with PACK over PART achieve a significantly lower overhead than AODV. Whenever route errors occur, AODV invokes the route discovery procedure and discovers a new route across the whole network.

\begin{figure}[H]
	\centering
	\includegraphics[scale=0.7]{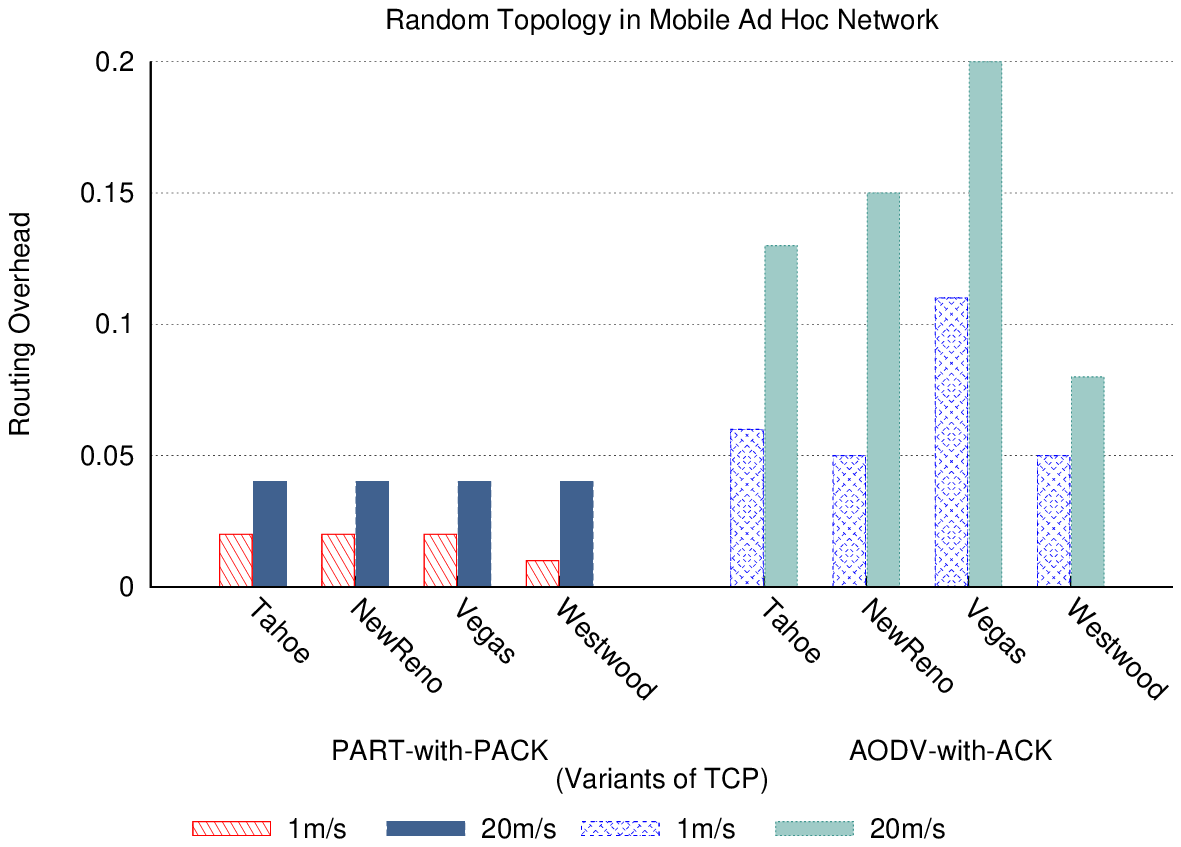}
	\caption{Routing overhead analysis across TCPs}
	\label{fig:overhead_random}
\end{figure}

On the contrary, in the PART protocol, if the route errors occur due to the packet collisions, PART protocol repairs the route locally instead of sending route error messages to the source node and broadcasting the request packets to the network in order to find a new route. These upshots affect PART protocol over variants of TCP with the proposed PACK mechanism. In Fig. 14, the PACK mechanism with TCP variants reduces routing overhead almost 66\% lower in Tahoe, 68\% lower in New Reno, 83\% lower in Vegas and 95\% lower in Westwood over PART routing protocol.

\subsubsection{Average delay measurements for variants of TCP} 
Fig. 15 shows the average delay measurement with TCP variants with different node speeds. The PACK mechanism over TCP variants suffers a higher delay at moderate speeds and starts achieving a lower delay starting from 20m/s speed. On average, the PACK mechanism reduces the average delay about 10\% lower in Tahoe, 8\% lower in Vegas and 9\% lower in Westwood when compared to AODV. When the node speeds increase, the possibilities of route break happen more often. The PART routing protocol recovers the route breaks using proxy nodes, which will result in lower delay.

\begin{figure}[!h]
	\centering
	\includegraphics[scale=0.7]{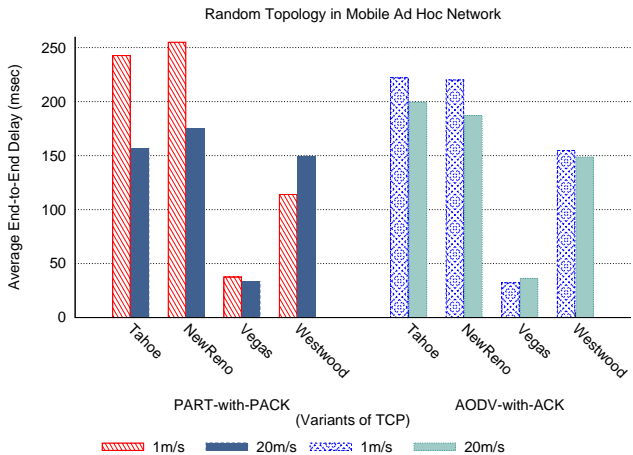}
	\caption{Average delay analysis across TCPs}
	\label{fig:delay_random}
\end{figure}

\section{\uppercase{Conclusion}}
We present a new mechanism called proxy acknowledgement (PACK) that detects any missing sequence numbers between a pair of source and destination. The PACK mechanism is applied to the variants of TCP. The performance differences are analyzed by varying network topologies in static and mobile ad hoc environments. Simulation results show that PACK provides better performance with TCP variants over PART protocol. In the grid topology, PACK over PART with TCP variants provides a better throughput and a lower packet loss up to 60\%. In random topology, the node speeds are varied at 1m/s and 20m/s node speed. Simulation results show that PACK with TCP variants has a higher throughput up to 11\%, a lower delay up to 10\% and a lower routing overhead up to 95\% over PART.

\bibliographystyle{jcn}


\begin{thebibliography}{10}
	
	\bibitem{Postel81}
	J. Postel, ``Transmission Control Protocol (TCP),'' {\em Request for Comments 793},
	1981.
	
	\bibitem{Allman02}
	M. Allman and S. Floyd, ``Increasing TCP's Initial Window,'' {\em Request for Comments 3390}, 2002.
	
	\bibitem{Stevens01}
	W. Stevens, ``TCP Slow Start, Congestion Avoidance, Fast Retransmit,'' in {\em Request for Comments 2001}, 1997.
	
	\bibitem{Chen08}
	J. Chen, Y.~Z. Lee, M. Gerla and M.~Y. Sandidi, ``TCP with delayed ack for wireless networks,'' {\em Ad Hoc Network}, vol.~6, pp.~1098--1116, 2008.
	
	\bibitem{Allman99}
	M. Allman, ``TCP Congestion Control,'' {\em Request for Comments 2581}, 1999.
	
	\bibitem{Floyd99}
	S. Floyd and T. Henderson, ``The NewReno Modification to TCP's Fast Recovery Algorithm,'' {\em Request for Comments 2582}, 1999.
 
	\bibitem{Mathis96}
	M. Mathis, J. Mahdavi and S. Floyd, ``TCP Selective Acknowledgement Options,'' {\em Request for Comments 2018}, 1996.
	
	\bibitem{Allman99}
	L.~S. Brakmo, S.!W. O'Malley and L.~L. Peterson, ``TCP Vegas: new techniques for congestion detection and avoidance,'' in {\em ACM SIGCOMM}, 1994.
	  
	\bibitem{Casetti02}
	 C. Casetti, M. Gerla, S. Mascolo, M.~Y. Sanadidi and R. Wang, ``TCP Westwood: End-to-End Congestion Control for Wired/Wireless Networks,'' in {\em Wireless Networks}, vol.~8, pp.~467--479, 2002.
    
    \bibitem{Hampel13}
    G. Hampel, A. Rana and T. Klein, ``Seamless TCP Mobility using Lightweight MPTCP Proxy,'' in {\em ACM MobiWac}, 2013.

	 \bibitem{Scofield08}
	 D. Scofield, L. Wang and D. Zappala, ``HxH: a hop-by-hop transport protocol for multi-hop wireless networks,'' in {\em the 4th Annual Int. Conf. on Wireless Internet}, 2008.

	\bibitem{Heimlicher07}
	S. Heimlicher, R. Baumann, M. May and B. Plattner, ``The Transport Layer Revisited,'' in {\em the 2nd Int. Conf. on Communication Systems, Software and Middleware}, 2007.

	\bibitem{Kopprty02}
	S. Kopparty, S.~V. Krishnamurthy, M. Faloutsos and S.~K. Tripathi, ``Split TCP for mobile ad hoc networks,'' in {\em IEEE GLOBECOM}, 2002.
	
	\bibitem{Oo11}
	M.~Z. Oo and  M. Othman, ``A proxy-assisted routing for efficient data transmission in mobile ad hoc networks,'' in {\em Wireless Networks}, vol.~17, pp.~1821--1832, 2011. 
	
	\bibitem{ns2}
	``{NS-2} project.'' http://www.isi.edu/nsnam/ns/, 2015. 
		
\end{thebibliography}

\newpage
\epsfysize=3.2cm
\begin{biography}{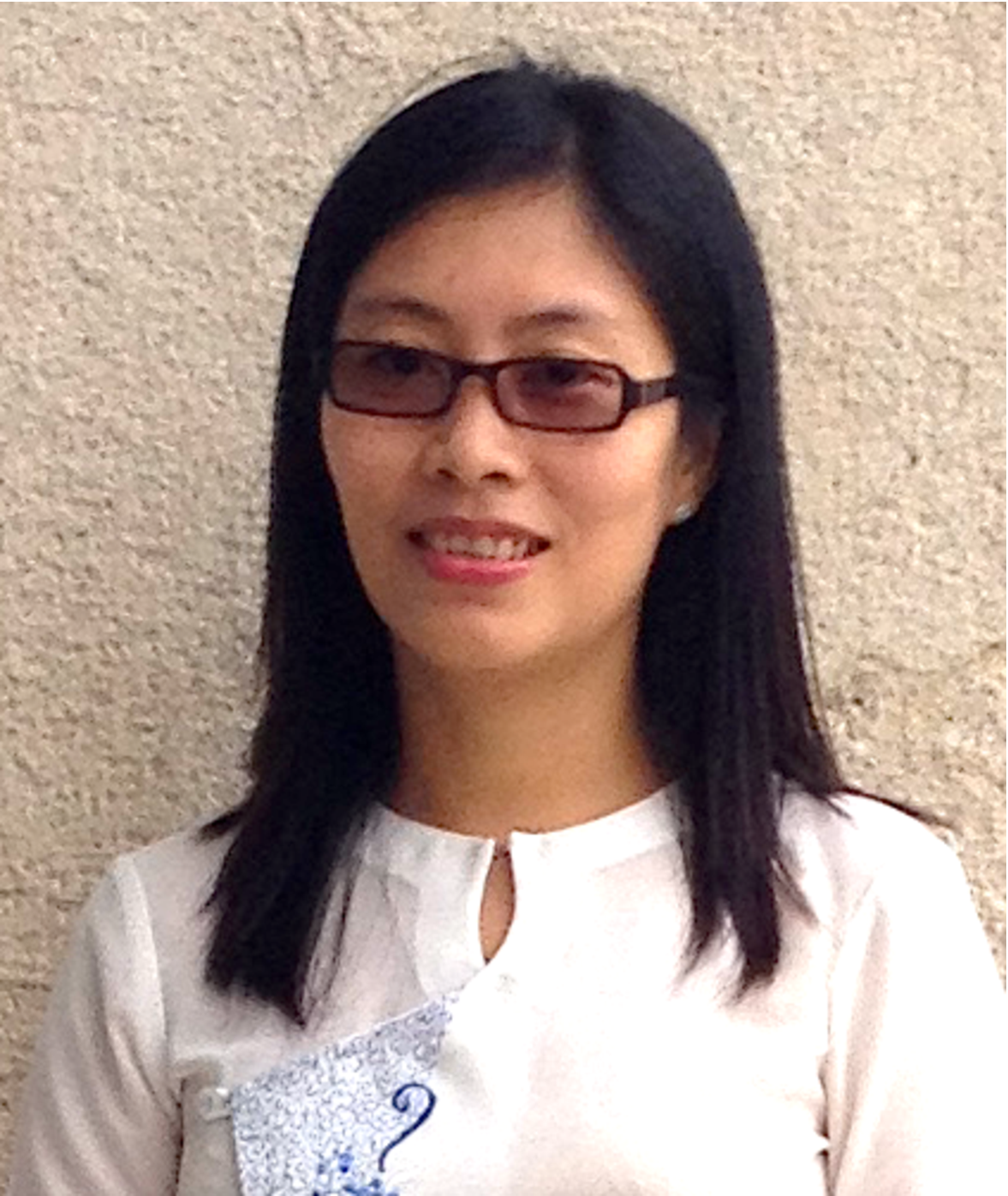}{May Zin Oo} is a lecturer at Mandalay Technological University, Myanmar. She received her first B.E. degree from Mandalay Technological University. Later she received her M.E. degree from  Yangon Technological University. Afterwards, she obtained a Ph.D. degree in wireless communication from the University of Malaya, Malaysia in 2012.  Her research interests include wireless communication and networking. \\
\end{biography}

\epsfysize=3.2cm
\begin{biography}{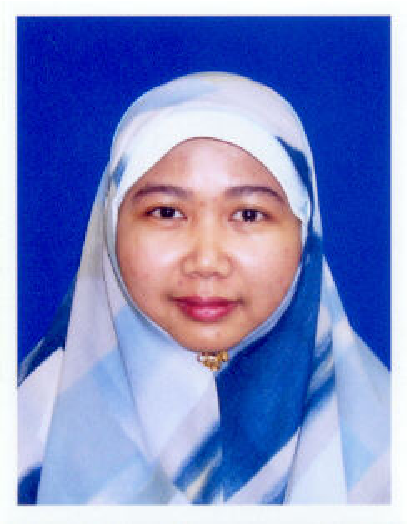}{Mazliza Othman} is an associate professor with the Faculty of Computer Science and IT at the University of Malaya, Malaysia. She received her B.Sc. in Computer Science from Universiti Kebangsaan Malaysia. Later she obtained a M.Sc. in Data Communication Networks and Distributed Systems and Ph.D. degree in mobile computing from the University College London, UK.  Her research interests include pervasive computing and self-organizing networks. She is the author of 'Principle of Mobile Computing and Communications.'\\
	
\end{biography}

\epsfysize=3.2cm
\begin{biography}{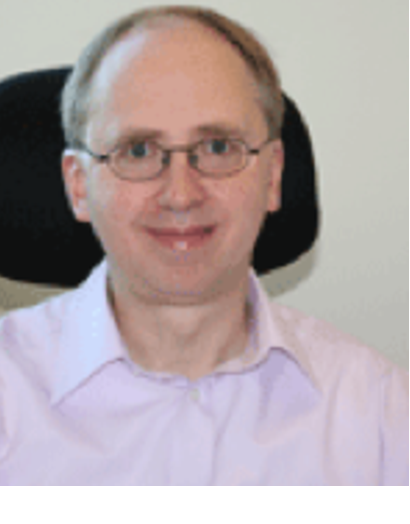}{Timothy O'Farrell} is Chair Professor of Wireless Communications at the University of Sheffield, UK. He is an expert in wireless communication systems specialising in physical layer signal processing, radio resource management and wireless network planning. He has pioneered research on energy efficient mobile cellular communications, the mathematical modeling of CSMA based MAC protocols for WiFi, iterative block coding for wireless communication systems and spreading sequence design for CDMA wireless networks. He is an entrepreneur, being the cofounder and CTO of Supergold Communication (2000-2007), a start-up that participated in the standardisation of IEEE 802.11g with the MBCK proposal. In the framework of Mobile VCE (mVCE), Professor O'Farrell was the Academic Coordinator of the Core 5 Green Radio project (2009-2012) and a leader in establishing energy efficiency as a global research field in wireless communication systems. He has managed 20 major research projects as principle investigator; published over 250 journal and conference papers; published 15 patents; and has participated in standards, consultancies and expert witness activities within the wireless sector. Currently, Professor O'Farrell leads the UK Research Strategy Community Organisation in Communications, Mobile Computing and Networking funded by EPSRC. Professor O'Farrell is a Chartered Engineer and a member of the IET and IEEE.
\end{biography}

\end{document}